\documentclass{article}
\usepackage[utf8]{inputenc}
\usepackage{authblk}

\usepackage[english]{babel}
\usepackage{graphicx}
\usepackage{txfonts}

\usepackage{amssymb}	
\usepackage{textcomp}
\usepackage{gensymb}
\usepackage{changes}

\usepackage{aas_macros}

\title{On the nature of the 35-day cycle in HZ~Her/Her~X-1\footnote{Paper presented at the Fourth Zeldovich virtual meeting, September 7-11, 2020.}}

\author[1,2]{N. I. Shakura} 
\author[1]{D. A. Kolesnikov} 
\author[1,2]{K. A. Postnov}

\affil[1]{Sternberg Astronomical Institute, Moscow State University, 119234 Moscow, Russia}
\affil[2]{Kazan Federal University, 420008 Kazan, Russia}

\begin{document}
\maketitle

\begin{abstract}
    Regular variations of the pulse period of Her X-1 with X-ray flux observed by \textit{Fermi}/GBM are examined. We argue that these regular variations result from free precession of the neutron star in Her X-1.
\end{abstract}

Among outstanding discoveries of the 1960s, the most bright is the discovery of accreting black holes and neutron stars in close binary stars made by the \textit{Uhuru} X-ray satellite. Her X-1 is one of the first X-ray pulsars discovered. It is a magnetized accreting neutron star (NS) in a 1.7-day orbit with the Roche-lobe filling optical star HZ Her  \cite{1972ApJ...174L.143T}. The NS mass is $m_{\rm x} = 1.4\,M_{\odot}$, the NS spin period is $P_{\rm x} = 1.24$~s. The mass of HZ Her is $m_{\rm o} = 2.0\,M_{\odot}$.

The optical brightness of HZ Her demonstrates a significant modulation with the orbital  period of $P_\mathrm{b}=1.7$ d. This modulation is due to a strong irradiation effect of the donor star \cite{1972IBVS..720....1C, 1972ApJ...178L...1B}. The X-ray light curve of Her X-1 shows sharp eclipses because of the high inclination of the binary system to the line of sight, about $90\degree$.

In X-ray binaries similar to HZ Her/Her X-1, the mass transfer from the optical component to the compact star occurs  through the inner Lagrangian point to form a turbulent near-Keplerian accretion disk around the compact object.
Due to turbulent viscosity, the matter in the disk loses the angular momentum and slowly approaches the central object.
The heat released during accretion is radiated from the disk surface. In Her X-1, at a distance of $\sim 100\,R_{\rm NS}$ from the center, the NS magnetic field breaks the disk, and at lower distances the matter freely falls along the magnetic field lines towards the NS surface and stops near the magnetic poles. Near the surface, the velocity of the infalling matter is close to one third of the speed of light.
During the collision, the huge kinetic energy of the accreting plasma is transformed into heat and is radiated away in X-rays.    

Soon after the discovery of the X-ray source in 1971 \cite{1972ApJ...174L.143T}, a 35-d  modulation of the X-ray flux from Her X-1 was discovered by the \textit{Uhuru} satellite \cite{1973ApJ...184..227G}. 
The 35-day X-ray cycle of Her X-1 comprises four states: (1) the Main-on lasting for $\sim 7$ orbits with the highest X-ray flux;  (2) the first low state lasting for $\sim 4$ orbits; (3) the Short-on lasting  for $\sim 4$ orbits with the X-ray flux about three times as low as in the Main-on; (4) the second low-state lasting for $\sim 4$ orbits (see, e.g., \cite{1998MNRAS.300..992S} for more detail and \cite{2020ApJ...902..146L} for the recent update). 

The 35-d X-ray cycle is associated with a tilted, retrograde precessing accretion disk. In the middle of the Main-on and Short-on, the disk is maximum open to the observer, and the central X-ray source is visible. During the low states, the outer parts of the tilted disk block the X-ray source. Recent observations by the \textit{SRG}/eROSITA telescope during the low states of Her X-1 revealed the orbital variations of the X-ray flux \cite{2021..A&A..Shakura..et..al}. The analysis of a large amount of the optical photometric data of HZ Her \cite{1973ApJ...186..617B, 1978pans.proc..111B} independently supports the presence of such a disk.

On the other hand, the model of free precession of the neutron star has been suggested as the possible explanation of 35-d X-ray cycle \cite{1972Natur.239..325B, 1973AZh....50..459N,     1973NPhS..246...87K}. The \textit{EXOSAT} observations showed that the X-ray pulse profiles of Her X-1 change during the 35-d cycle \cite{1986ApJ...300L..63T}.  Later on, the \textit{Ginga} and \textit{RXTE} observations were used to study in detail the evolution of X-ray pulses with the 35-day phase  \cite{1998ApJ...502..802D,2000ApJ...539..392S, 2013A&A...550A.110S}. 

The \textit{RXTE} observations of Her X-1 suggest that the observed evolution of X-ray pulses could be explained by free precession of a NS with complex surface magnetic field \cite{2013MNRAS.435.1147P}. The NS free precession is also able to explain the optical light curves of HZ Her \cite{Shakura_ufn_2019, 2020MNRAS.499.1747K}. In this model, a more stable 35-day NS free precession period serves as a clock mechanism of the entire 35-day cycle via synchronization of the disk precession period by the action of gas streams forming the outer parts of the disk \cite{1999A&A...348..917S}. Possible synchronization mechanisms are further discussed in \cite{Shakura_ufn_2019, 2020MNRAS.499.1747K}.

As shown for the first time in \cite{1995pns..book...55S}, the NS free precession should be accompanied by small (about several microseconds) regular variations of the observed pulse period. These variations are clearly detected  in the \textit{Fermi}/GBM  data\footnote{https://gammaray.nsstc.nasa.gov/gbm/science/pulsars/lightcurves/herx1.html}. 

\begin{figure}
    \centering
    \includegraphics[width=\columnwidth]{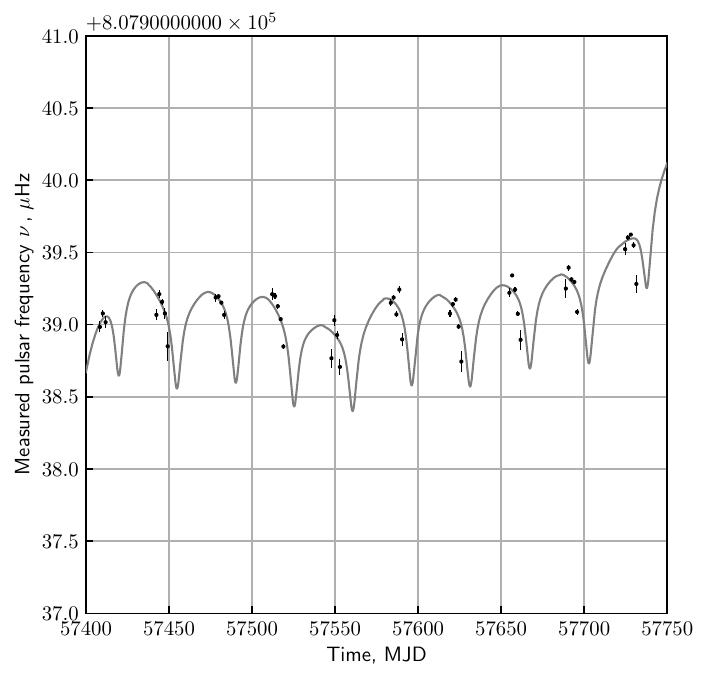}
    \caption{The measured pulsar frequency of Her X-1 as a function of time. Black dots show the \textit{Fermi}/GBM measurements. Only points with errors smaller than 0.1 $\mu$Hz are shown. The solid line indicates theoretical variation of the pulsar frequency due to NS free precession, Eq. \ref{eq:omega_m}. The slow drift of the pulsar frequency on timescale longer than 35 d can be due to possible irregular variations in the NS free precession parameters.}.
    \label{fig:variations}
\end{figure}

The observable pulse frequency is a sum of non-periodical and periodical variations:
\begin{equation}
    \omega_{\rm o}(t) = \omega_{\rm ns}(t) + \frac{d\phi(t)}{dt}
    \label{eq:omega_m}
\end{equation}
\begin{figure}
    \centering
    \includegraphics{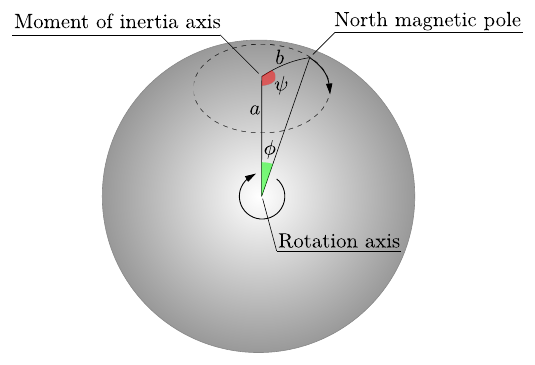}
    \caption{Scheme of the NS  free precession. The NS rotational equator is in the picture plane, the NS spin axis (R) is perpendicular to the figure plane. The dashed line indicates the path of the north magnetic pole (N) path of the freely precessing NS. The center of the dashed circle coincides with the NS principal momentum of inertia (I). The sides of the spherical triangle RNI are $a = 50\degree$ and $b = 30\degree$. $\psi$ is the free precession angle. The time derivative of the angle $\phi$ defines the pulse frequency variation.}
    \label{fig:free_precession_scheme}
\end{figure}
As seen from  Fig.\ref{fig:free_precession_scheme}, the NS free precession should modulate the observed pulse frequency. 
The angle $\phi$ can be found from sine and cosine theorem for spherical triangles:
\begin{equation}
  \cos\phi(t) = \frac{\sin{b}\sin{\psi(t)}}{\sqrt{1 - [\cos{a}\cos{b} + \sin{a} \sin{b} \cos{\psi(t)}]^2}}
\end{equation}
Here $a=50\degree$ is the side of the spherical triangle connecting the NS spin and inertia axes, $b$ is the side of the  spherical triangle connecting the NS inertia axis and the magnetic pole, $\psi(t)$ is the NS free precession angle  linearly depending on time: 
\begin{equation}
    \psi(t) = \Omega t + \psi_0
\end{equation}
Here $\Omega$ is the NS free precession angular frequency. The function $\omega_{\rm o}(t)$ is shown in Fig. \ref{fig:variations}. 

However, the \textit{Fermi}/GBM pulse frequency measurements show high-amplitude irregular variations. Fig. \ref{fig:variations} shows that the observed \textit{Fermi}/GBM variations agree with theoretical behaviour of $\omega_{\rm o}(t)$. 

So far we have supposed only two-axial  NS free precession. However, in Her X-1 a triaxial NS free precession is also possible \cite{1998A&A...331L..37S}. 

The authors acknowledge Drs. M. R. Gilfanov and S. V. Molkov for drawing attention to the \textit{Fermi}/GBM  observations of Her X-1. The research is supported by the RFBR grant no. 18-502-12025 and the Interdisciplinary Scientific and Educational School of Moscow University ``Fundamental and Applied Space Research''.

\bibliographystyle{ieeetr}

\bibliography{bib}

\end{document}